\documentclass[lettersize,journal]{IEEEtran}
\usepackage{amsmath,amsfonts}
\usepackage{algorithmic}
\usepackage{algorithm}
\usepackage{array}
\usepackage{textcomp}
\usepackage{stfloats}
\usepackage{url}
\usepackage{verbatim}
\usepackage{graphicx}
\usepackage{amssymb} 
\usepackage{amsfonts}
\usepackage{amsmath} 
\usepackage{physics} 
\usepackage{algorithmic}  
\usepackage{algorithm} 
\usepackage{graphicx}
\usepackage{color} 
\usepackage{bm}
\usepackage{epstopdf} 
\usepackage{subfigure}
\usepackage{booktabs}  
\usepackage[dvipsnames]{xcolor} 
\usepackage{cite} 
\usepackage{balance}
\usepackage{setspace}  
\usepackage{amsthm}
\usepackage{amsmath} 
\usepackage{hyperref}

\newcounter{MYtempeqncnt}
\def\widebreve{\mathpalette\wide@breve}
\def\wide@breve#1#2{\sbox\z@{$#1#2$}%
	\mathop{\vbox{\m@th\ialign{##\crcr
				\kern0.08em\brevefill#1{0.8\wd\z@}\crcr\noalign{\nointerlineskip}%
				$\hss#1#2\hss$\crcr}}}\limits}
\def\brevefill#1#2{$\m@th\sbox\tw@{$#1($}%
	\hss\resizebox{#2}{\wd\tw@}{\rotatebox[origin=c]{90}{\upshape(}}\hss$}
\makeatletter

\begin{document}

	\author{Yafei Wang, \textit{Graduate Student Member}, \textit{IEEE}, Hongwei Hou, \textit{Graduate Student Member}, \textit{IEEE},\\ Wenjin Wang, \textit{Member}, \textit{IEEE}, Xinping Yi, \textit{Member}, \textit{IEEE}
	\thanks{Manuscript received xxx.}
		\thanks{Yafei Wang, Hongwei Hou, Wenjin Wang, and Xinping Yi are with the National Mobile Communications Research Laboratory, Southeast University, Nanjing 210096, China (e-mail: wangyf@seu.edu.cn; 	
		hongweihou@seu.edu.cn; wangwj@seu.edu.cn; xyi@seu.edu.cn).}}
		%
		\title{Symbol-Level Precoding for Average SER Minimization in Multiuser MISO Systems}
		%
		%
		%

		%
		%

	\markboth{}%
	{Shell \MakeLowercase{\textit{et al.}}: Bare Demo of IEEEtran.cls for IEEE Journals}
	%



	\maketitle

	\begin{abstract}
		This paper investigates symbol-level precoding (SLP) for high-order quadrature amplitude modulation (QAM) aimed at minimizing the average symbol error rate (SER), leveraging both constructive interference (CI) and noise power to gain superiority in full signal-to-noise ratio (SNR) ranges. We first construct the SER expression with respect to the transmitted signal and the rescaling factor, based on which the problem of average SER minimization subject to total transmit power constraint is further formulated.
		Given the non-convex nature of the objective, solving the above problem becomes challenging.
	 	Due to the differences in constraints between the transmit signal and the rescaling factor, we propose the double-space alternating optimization (DSAO) algorithm to optimize the two variables on orthogonal Stiefel manifold and Euclidean spaces, respectively. 
		To facilitate QAM demodulation instead of affording impractical signaling overhead, we further develop a block transmission scheme to keep the rescaling factor constant within a block. 
		Simulation results demonstrate that the proposed SLP scheme exhibits a significant performance advantage over existing state-of-the-art SLP schemes.
	\end{abstract}
	

	%
	\IEEEpeerreviewmaketitle

	\section{Introduction}
	%
	%
	%
	%
	\IEEEPARstart{I}{n} 
	MULTIUSER multi-input multi-output (MU-MIMO) transmission, precoding techniques are applied to suppress interference among users and enhance spectral efficiency. Linear precoding schemes, exemplified by maximum ratio transmission (MRT) \cite{1468466}, zero-forcing (ZF), and minimum mean squared error (MMSE) \cite{1261332,1391204}, design the precoding matrix utilizing channel state information (CSI) with low computational complexity.
	In contrast, nonlinear precoding further improves the performance by exploiting the information of input data, e.g., Tomlinson-Harashima precoding (THP) \cite{1091221}, dirty paper coding (DPC) \cite{1056659}, and symbol-level precoding (SLP) \cite{8359237, 9035662}.
	
	Among nonlinear precoding, SLP leverages interference with the user symbols and their corresponding constellations to design precoding schemes at a symbol level \cite{8359237, 9035662, Li2021,4801492,5605266,7103338,7942010,8477154,8374931,9120670,shao2020minimum}. 
	In \cite{5605266}, the idea of transforming destructive interference (DI) to constructive interference (CI) by rotating phase was considered, which is the embryonic form of SLP. Since SLP is usually designed for uncoded systems, symbol error rate (SER) minimization becomes a crucial criterion for SLP design \cite{8374931, 9120670, shao2020minimum,9054488}. 
	 In \cite{8374931}, the close form of SER for phase-shift keying (PSK) was analyzed,  and the SER minimization problem was investigated with CI. To minimize the sum-SER of PSK, a power allocation scheme for SLP transmission was proposed in \cite{9120670}. In \cite{9054488}, a deep learning transceiver was constructed to reduce SER in quadrature amplitude modulation (QAM) transmission. Furthermore, SLP was jointly designed with intelligent
	 reflecting surface (IRS) to minimize the SER for PSK and QAM in \cite{shao2020minimum}. 
	 
	 Most existing SLP schemes could only promise significant performance gains in the high signal-to-noise ratio (SNR) regime since they control the interference but ignore the impact of the noise. By further utilizing the noise power, SLP can achieve better performance in low SNR regimes \cite{9910472}. Based on this observation, this paper aims to design SLP with average SER minimization for high-order QAM transmission, which exploits both the constellations and the noise power to gain superiority over the full SNR region. Firstly, we construct SER expression, based on which the problem of average SER minimization subject to total transmit power constraint is further formulated. Given the non-convex nature of the objective, solving the above problem becomes challenging.
	 Due to the differences in constraints between the transmit signal and the rescaling factor, we propose the double-space alternating optimization (DSAO) algorithm to optimize the two variables on orthogonal Stiefel manifold and Euclidean spaces, respectively. 
	 In order to facilitate QAM demodulation in the receiver, we further develop a block transmission scheme for the proposed SLP design. Finally, simulation results demonstrate that the proposed SLP scheme exhibits a significant performance advantage over existing state-of-the-art SLP schemes.
	
	\section{System Model}\label{S2}
	
	\subsection{System Model}
	Consider a downlink system that consists of one $N$-antenna base station (BS) and $K$ single-antenna user equipments (UEs), while it is assumed that $N\geq K$. We consider the block flat fading channels, where the channel coefficients remain constant during a coherence interval of $L$ symbol durations. The channel between BS and the $k$-th UE is denoted as  ${\bf h}_k\in{\mathbb{C}}^{N\times 1}$, and the CSI $\{{\bf h}_k\}_{k\in{\mathcal{ K}}}, {\mathcal{ K}}=\left\{1, 2, ..., K\right\}$ is assumed to be perfectly available at the BS.

	At the $l$-th symbol duration, $K$ independent QAM symbols are intended to be transmitted to $K$ UEs. The symbols are mapped to the transmit vector ${\bf x}[l]\in{\mathbb{C}}^{N\times 1}$ by symbol-level precoder ${\rm SLP}\left(\cdot\right)$, which can be expressed as
	\begin{align}
		{\bf x}[l] = {\rm SLP}\left(\{s_{k}[l]\}_{k\in{\mathcal{ K}}},\{{\bf h}_k\}_{k\in{\mathcal{ K}}}, \sigma^2\right),\ \forall l \in {\mathcal{ L}},
	\end{align}
	where $s_{k}[l]$ is the symbol belonging to the $k$-th UE, $\sigma^2$ represents the noise power, $n_k[l]\sim \mathcal{C}\mathcal{N}(0, \sigma^2)$ denotes the additive noise at the $k$-th UE, and ${\mathcal{ L}}=\{1, 2, ..., L\}$. The received signal of $k$-th UE is
	\begin{equation}
		{y}_k[l] = {\bf h}^T_k{\bf x}[l] + n_k[l].
		\label{E1}
	\end{equation}
	When multi-level modulations are employed, the received signals require to be scaled for correct demodulation, and the signal to be demodulated is
	\begin{align}
		\begin{split}
			\bar{y}_k[l] &= {{y}_k[l]}/{\gamma[l]}\\
			&= \left({{\bf h}^T_k{\bf x}[l] + n_k[l]}\right)/{\gamma[l]},
		\end{split}
	\end{align}
	where ${\gamma[l]}$ is the rescaling factor optimized by SLP scheme \cite{li2020symbol}.

	\begin{figure}[!t]
		\centering
		\includegraphics[width=3in]{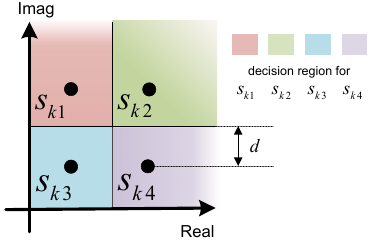}
		\caption{Decision region of 16QAM constellation points in the first quadrant.}
		\label{Decision region}
	\end{figure}
	
	\section{Symbol-Level Precoder Design for \\Average SER Minimization}\label{S3}		
	In this section, we design the SLP that minimizes average SER minimization (ASM) under the symbol-level transmit power constraint.
		\subsection{Problem Formulation}\label{asm problem}
			
			We first derive the SER expression for QAM symbols with the maximum likelihood (ML) decision rule. Due to the independence of $\real(n_k[l])$ and $\imaginary(n_k[l])$, the SER of ${\bar{y}}_k[l]$ corresponding to $s_k[l]$ can be obtained by
			\begin{align}
				E_{k}[l] = 1-\left(1-E^{\rm R}_{k}[l]\right)\cdot \left(1-E^{\rm I}_{k}[l]\right), 
				\label{SER2}
			\end{align}
			where $E^{\rm R}_{k}[l]$ and $E^{\rm I}_{k}[l]$ denote the error rate of $\real({\bar{y}}_k[l])$ and $\imaginary({\bar{y}}_k[l])$, respectively.
			
			\begin{figure*}[!hb]
				\vspace*{4pt}
				\hrulefill
				\begin{align}
				E^{\rm R}_{k({\rm i})}[l] \!=\!
				1\!-\! {\rm Pr}\left({\hat s}^{\rm R}_{k}[l]+d> \frac{\real({\hat {\bf h}}^{\rm T}_k[l]{\bf x}[l]\!+\!{\hat n}_k[l])}{\gamma[l]}>{\hat s}^{\rm R}_{k}[l]\!-\! d\right),\ 		E^{\rm R}_{k({\rm o})}[l] \!=\!  1\!-\! {\rm Pr}\left(\frac{\real({\hat {\bf h}}^{\rm T}_k[l]{\bf x}[l]\!+\!{\hat n}_k[l])}{\gamma[l]} >{\hat s}^{\rm R}_{k}[l]\!-\! d\right),
				\label{SER_R_1}
				\end{align}
			\end{figure*}
			
			For the simplicity of the subsequent representations, we introduce the following notations. To unify the SER expressions of QAM constellation points in all four quadrants, we define
			\begin{align}
			\begin{split}
			&\theta_k[l] = \angle\left(a^{\rm R}_{k}[l]+ja^{\rm 		I}_{k}[l]\right)-\frac{\pi}{2 },\ \hat{\bf h}_k \!=\!  {\bf h}_ke^{-j\theta_k[l]},\\
			&\hat{s}_k[l] \!=\! {s}_k[l]e^{-j\theta_k[l]},\  \hat{n}_k[l] \!=\! {n}_k[l]e^{-j\theta_k[l]},
			\end{split}\label{rotation}
			\end{align}
			where $j=\sqrt{-1}$ is the imaginary unit, $\angle(\cdot)$ denotes the function that extracts the angle in radians, and
			\begin{align}
			\begin{split}
			s^{\rm R}_{k}[l] = \real(s_{k}[l]),& \ s^{\rm I}_{k}[l] = \imaginary(s_{k}[l]), \\
			a^{\rm R}_{k}[l] = {\rm sign}\left(s^{\rm R}_{k}[l]\right),& \ a^{\rm I}_{k}[l] = {\rm sign}\left(s^{\rm I}_{k}[l]\right),
			\end{split}
			\end{align}
			where ${\rm sign}(\cdot)$ represents the sign function.
			Furthermore, we denote $ {\mathcal{ O}}$ as the set consisting of the outer real and imaginary parts of constellations (e.g., $s^{\rm I}_{k2}$, $s^{\rm R}_{k2}$, and $s^{\rm R}_{k4}$ in Fig. \ref{Decision region}), and $ {\mathcal{ I}}$ is the set consisting of the inner parts (e.g., $s^{\rm R}_{k3}$, $s^{\rm I}_{k3}$, and $s^{\rm R}_{k4}$) \cite{Li2021}. Besides, $E^{\rm R}_{k({\rm i})}[l]$ and $E^{\rm R}_{k({\rm o})}[l]$ represent SER corresponding to $s^{\rm R}_{k}[l]\in {\mathcal{ I}}$ and ${\mathcal{ O}}$, respectively.
		
			Based on the above notation definitions, $E^{\rm R}_{k}[l]$ in \eqref{SER2} can be expressed as \eqref{SER_R_1}, where ${\rm Pr}(\cdot)$ denotes probability, $d$ is half of the distance between adjacent constellations, ${\hat s}^{\rm R}_{k}[l]$ and ${\hat s}^{\rm I}_{k}[l]$ represent the real and imaginary parts of ${\hat s}_{k}[l]$. We denote the transmit power budget of ${\bf x}[l]$ as $P_{\rm T}[l]$ and introduce the subsequent definitions
					\begin{align}
						\begin{split}
						&{\bf f}_k[l] \!=\! \sqrt{P_{\rm T}[l]}  
						\begin{bmatrix}
							\real({\hat {\bf h}}_k[l])\\
							\!-\!\imaginary({\hat {\bf h}}_k[l])
						\end{bmatrix},
						{\bf g}_k[l] \!=\! \sqrt{P_{\rm T}[l]}
						\begin{bmatrix}
							\imaginary({\hat {\bf h}}_k[l])\\
							\real({\hat {\bf h}}_k[l])
						\end{bmatrix},\\
						&\ \ {\bar{\bf x}}[l] \!=\!  \frac{1}{\sqrt{P_{\rm T}[l]}}
						\begin{bmatrix}
							\real({\bf x}[l])\\
							\imaginary({\bf x}[l])
						\end{bmatrix},
						b^{\rm R}_{k}[l] \!=\!
						\begin{cases}
							1\ \ {\rm if}\  {s}^{\rm R}_{k}[l]\in{\mathcal{ I}}\\
							0\ \ {\rm if}\  {s}^{\rm R}_{k}[l]\in{\mathcal{ O}}
						\end{cases},
						\end{split}
						\label{real notation}
					\end{align}
					based on which we have $\real({\hat {\bf h}}^T_k[l]{\bf x}[l])={\bf f}^T_k{\bar {\bf x}}$ and $\imaginary({\hat {\bf h}}^T_k[l]{\bf x}[l])={\bf g}^T_k{\bar {\bf x}}$.
					With the distribution of $n_k[l]$, $E^{\rm R}_{k}[l]$ can be written as
					\begin{align}
						E^{\rm R}_{k}[l] = 1-Q\left(\frac{\gamma[l]{\hat s}^{\rm R}_{k}[l]-\gamma[l]\cdot d-{\bf f}^T_k[l]\bar{\bf x}[l]}{\sigma/\sqrt{2}}\right)\notag\\
						+b^{\rm R}_{k}[l]Q\left(\frac{\gamma[l]{\hat s}^{\rm R}_{k}[l]+\gamma[l]\cdot d-{\bf f}^T_k[l]\bar{\bf x}[l]}{\sigma/\sqrt{2}}\right),
						\label{SER1_real}
					\end{align}
					where $Q(x) = \int^{\infty }_{x}\frac{1}{\sqrt{2\pi}}e^{-\frac{t^2}{2}}{\rm d}t$. For the imaginary part, $b^{\rm I}_{k}[l]$ has a definition similar to $b^{\rm R}_{k}[l]$, and $E^{\rm I}_{k}[l]$ can be expressed similarly to $E^{\rm R}_{k}[l]$ by replacing $b^{\rm R}_{k}[l]$, ${s}^{\rm R}_{k}[l]$, and ${\bf f}_k[l]$ with $b^{\rm I}_{k}[l]$, ${s}^{\rm I}_{k}[l]$, and ${\bf g}_k[l]$, respectively.

					Based on (\ref{SER2}) and (\ref{SER1_real}), we formulate the optimization problem of ASM, which optimizes ${\bar{\bf x}}[l]$ and $\gamma[l]$ to minimize the average SER subject to transmit power budget
					\begin{align} 
					\begin{split}
					\min\limits_{{{\bar{\bf x}}[l]}, \gamma[l]}&\frac{1}{K}\sum_{k=1}^{K}E_k[l], \\{\rm s.t.}&~  \left \|{\bar{\bf x}}[l]\right \|_{2}^{2} \le 1,
					\end{split}
					\label{sum-SER}
					\end{align}	
					where $\left \|\cdot\right \|_{2}$ denotes $l_2$-norm and 
					\begin{align}
					E_k[l]=1-\left[Q\left(\frac{\gamma[l]\hat{s}^{\rm R}_{k}[l]-\gamma[l]\cdot d-{\bf f}^T_k[l]{\bar {\bf x}}[l]}{\sigma/\sqrt{2}}\right)\right.&\notag\\
					\left.-b^{\rm R}_{k}[l]Q\left(\frac{\gamma[l]\hat{s}^{\rm R}_k[l]+\gamma[l]\cdot d-{\bf f}^T_k[l]{\bar {\bf x}}[l]}{\sigma/\sqrt{2}}\right)\right]&\notag\\
					\cdot
					\left[Q\left(\frac{\gamma[l]\hat{s}^{\rm I}_{k}[l]-\gamma[l]\cdot d- {\bf g}^T_k[l]{\bar {\bf x}}[l]}{\sigma/\sqrt{2}}\right)\right.&\notag\\
					\left.-b^{\rm I}_{k}[l]Q\left(\frac{\gamma[l]\hat{s}^{\rm I}_{ k}[l]+\gamma[l]\cdot d- {\bf g}^T_k[l]{\bar {\bf x}}[l]}{\sigma/\sqrt{2}}\right)\right].
					\label{SER expression}
					\end{align}
		
		\subsection{Double Space Alternating Optimization Algorithm}\label{ao algorithm}
		Due to the non-convexity of the cost function, \eqref{sum-SER} is difficult to solve by conventional gradient descent-type algorithms. Due to the differences in constraints between the transmit signal ${\bar{\bf x}}[l]$ and the rescaling factor $\gamma[l]$, we propose to optimize the two variables alternately.
		
		In the first alternating
		step, we optimize ${\bar{\bf x}}[l]$ with a
		fixed rescaling factor $\gamma[l]$, and problem \eqref{sum-SER} can be restated as
		\begin{align}
			\begin{split}
			&\ \ \min\limits_{{\bar{\bf x}}[l]}\ \ g\left(\bar{\bf x}[l]\right)\\
			&{\rm s.t.}~  \left \|\bar{\bf x}[l]\right \|_{2}^{2} = 1,
			\end{split}\label{problem_x2_1}
		\end{align}
		where we use $g(\cdot)$ to denote the cost function of \eqref{sum-SER} and the constraint $\left \|{\bar{\bf x}}[l]\right \|_{2}^{2} \le 1$ is replaced by $\left \|{\bar{\bf x}}[l]\right \|_{2}^{2} = 1$ to facilitate the subsequent solution.
		The constraint of the subproblem is a unit sphere $S^{2N-1}$ known as the orthogonal Stiefel manifold (OSM), defined as 
		\begin{align}
		{\mathcal{M}}(p, n) = \left\{{\bf X}\in\mathbb{R}^{n\times p}:{\bf X}^T{\bf X}={\bf I}_p\right\}.
		\end{align}
		Since the negative Riemannian gradient in manifold optimization (MO) is the steepest descend direction at the current point within the manifold \cite{P.A.Absil2009}, we apply its gradient descent algorithm to this problem.
		The update step of gradient descent in MO is given by \cite{P.A.Absil2009}
		\begin{align}
			{\bar{\bf x}}^{(m+1)}[l] = R_{\bar{\bf x}}\left(-t^{(m)}\nabla_{\mathcal{M}}g\left({\bar{\bf x}}^{(m)}[l]\right)\right),
			\label{x2 iteration}
		\end{align}
		where ${\bar{\bf x}}^{(m)}[l]$ denotes the $\bar{\bf x}[l]$ of the $m$-th iteration; $\nabla_{\mathcal{M}} g({\bar{\bf x}})={\rm P}_{{\bar{\bf x}}}(\nabla g({\bar{\bf x}}))$ is the Riemannian gradient, $\nabla g({\bar{\bf x}})$ is the Euclidean gradient of $g({\bar{\bf x}})$ with respect to ${\bar{\bf x}}$, and ${\rm P}_{{\bar{\bf x}}}({\boldsymbol{\xi}})={\boldsymbol{\xi}}-{\bar{\bf x}}{\bar{\bf x}}^T{\boldsymbol{\xi}}$ is the projection operator; $t^{(m)}$ is the chosen Armijo step size \cite{P.A.Absil2009}; $R_{\bar{\bf x}}$ represents the retraction function remaps the updated variables into the manifold, and the commonly used operator of $S^{2N-1}$ is
		\begin{align}
			R_{{\bar{\bf x}}}({\boldsymbol{\xi}})&=\frac{{\bar{\bf x}}+{\boldsymbol{\xi}}}{||{\bar{\bf x}}+{\boldsymbol{\xi}}||_2}.\label{shrinkage operators}
		\end{align}

		In the next alternating step, we fix ${\bar{\bf x}}[l]$ and find the best-matched
$\gamma[l]$
		\begin{align}
			\min\limits_{{\gamma}[l]}&\ \ g\left(\gamma[l]\right),
			\label{gamma problem}
		\end{align}
		which is a simple unconstrained optimization problem with the one-dimensional variable $\gamma[l]$. The gradient descent method is performed for iteration
		\begin{align}
			\gamma^{(m+1)}[l] = \gamma^{(m)}[l]-{\tau^{(m)}}[l]\nabla g(\gamma^{(m)}[l]),
			\label{gamma iteration}
		\end{align}
		where ${\tau^{(m)}}[l]$ is the step size choosed by backtracking line search \cite{Boyd2004}.
		Combining the iterative formulas \eqref{x2 iteration} and \eqref{gamma iteration}, the proposed DSAO algorithm is given in \textbf{Algorithm \ref{A1}}, where index $l$ is temporarily omitted for readability. The detail expressions of $\nabla g({\bar{\bf x}})$ and $\nabla g(\gamma)$ can be found in Appendix \ref{Gradient}.
		\begin{algorithm}[htbp]
			\caption{Double Space Alternating Optimization Algorithm for Problem \eqref{sum-SER}}
			\label{Heuristic DL-MMSE-DUAL}
			\begin{spacing}{1.1}
				\begin{algorithmic}[1]
					\STATE \textbf{Input:} $\{s_{k}\}_{k\in{\mathcal{ K}}}$, $\{{\bf h}_k\}_{k\in{\mathcal{ K}}}$, $\sigma$, $d$, $P_{\rm T}$, $\gamma^{(0)}$, $\bar{\bf x}^{(0)}$.
					\STATE Obtain $\left\{{\bf f}_k,{\bf g}_k,\hat{s}^{\rm R}_k,\hat{s}^{\rm I}_k,b^{\rm R}_k,b^{\rm I}_k\right\}_{k\in{\mathcal{ K}}}$ based on \eqref{rotation} and \eqref{real notation}.
					\STATE Initialize $m=0$.
					\STATE \textbf{repmat}
					\STATE  \quad Compute $\nabla g({\bar{\bf x}}^{(m)})$ based on \eqref{SER expression}.
					\STATE  \quad $\nabla_{\mathcal{M}} g({\bar{\bf x}}^{(m)})=\nabla g({\bar{\bf x}}^{(m)})-{\bar{\bf x}}^{(m)}({{\bar{\bf x}}^{(m)}})^T\nabla g({\bar{\bf x}}^{(m)})$ .
					\STATE  \quad Choose step size $t^{(m)}$ (\cite{P.A.Absil2009}, see section 4.2.2).
					\STATE  \quad ${\bar{\bf x}}^{(m+1)} = R_{\bar{\bf x}}\left(-t^{(m)}\nabla_{\mathcal{M}}g\left({\bar{\bf x}}^{(m)}\right)\right)$.
					\STATE  \quad Choose  ${\tau^{(m)}}$ by backtracking line search \cite{Boyd2004}.
					\STATE \quad $\gamma^{(m+1)} = \gamma^{(m)}-{\tau^{(m)}}\nabla
					g(\gamma^{(m)})$.
					\STATE  \quad $m = m+1$.
					\STATE \textbf{until} some stopping criterion is met.
					\STATE \textbf{Output:} ${\bar{\bf x}}^{(m)}$, $\gamma^{(m)}$.
				\end{algorithmic}\label{A1}
			\end{spacing}
		\end{algorithm}
		
		The computational complexity per iteration of \textbf{Algorithm \ref{A1}} is primarily determined by the computations of the average SER and related gradients, which has complexity order of $\mathcal{O}(N(N+K))$.
		
		\subsection{Block Transmission Scheme}\label{S5}
		
		When high-order QAM is used, the presence of $\gamma[l]$ becomes necessary at the UE to rescale the received signal for correct demodulation. Although most prior SLP designs have assumed that $\gamma[l]$ is known to the UEs, the practical scenario of block transmission necessitates the BS to send $\gamma[l]$ to the UEs at the symbol level. To prevent the introduced significant increase in signaling overhead, we apply the power allocation scheme proposed in \cite{li2020symbol} to unify $\gamma[l]$ in a block:
		\begin{equation}
			\begin{split}
				&\qquad\qquad\qquad\gamma_{\rm blk} = \sqrt{\frac{\sum_{l=1}^{L}P_{\rm T}[l]}{{\sum_{l=1}^{L}\frac{P_{\rm T}[l]}{\gamma^2[l]}}}},\\
				&{\bar P}_{\rm T}[l] = \left( \frac{\gamma_{\rm blk}}{\gamma[l]}\right)^2 P_{\rm T}[l],\ {\bar{\bf x}}_{\rm pa}[l] = \frac{\gamma_{\rm blk}}{\gamma[l]}{\bar{\bf x}}[l],\ \forall l \in {\mathcal{ L}},
			\end{split}\label{PA}
		\end{equation}
		where $\gamma_{\rm blk}$ represents the unified $\gamma$ of the entire block, ${\bar{\bf x}}_{\rm pa}[l]$ and ${\bar P}_{\rm T}[l]$ represent the transmitted signal (real representation) and transmit power in $l$-th symbol duration after power allocation. 
		
		When \textbf{Algorithm \ref{A1}} is employed in the low SNR range and the diagonal symbols in ${\bf s}[l]$ account for a large proportion, it may prompt $\gamma[l]\to 0$ to reduce the error of the diagonal symbols while sacrificing the rest symbols. According to \eqref{PA}, this will cause $\gamma_{\rm blk}\to 0$, resulting in a sharp increase of SER within the block. Thus, we set initial $\gamma^{(0)}[l]$ as the lower bound for $\gamma[l]$, and $\gamma$ will not be updated if $\gamma^{(m+1)}[l]<\gamma^{(0)}[l]$ in step 10 of \textbf{Algorithm \ref{A1}}.
		Besides, we further optimize ${\bar{\bf x}}_{\rm pa}[l]$ by performing steps 5-8 in \textbf{Algorithm \ref{A1}} after power allocation for keeping the optimality of $\bar{\bf x}_{\rm pa}[l]$ given $\gamma_{\rm blk}$ and ${\bar P}_{\rm T}[l]$. The entire process is summarized in \textbf{Algorithm \ref{block transmission}}.
		
		\begin{algorithm}[t]
			\caption{Block Transmission Scheme of ASM}
			\label{A2}
			\begin{algorithmic}[1]
				\STATE \textbf{Input:} $\sigma$, $d$, $\{{\bf h}_k\}_{k\in{\mathcal{ K}}}$, $\left\{\left\{s_{k}[l]\right\}_{k\in{\mathcal{ K}}},P_{\rm T}[l]\right\}_{l \in {\mathcal{ L}}}$.
				\STATE \textbf{for} $l=1$ \textbf{to} $L$ \textbf{do}
				\STATE \quad Solve (\ref{sum-SER}) for ${\bf s}[l]$ using Algorithm \ref{A1} $\Rightarrow {\bar{\bf x}}[l], \gamma[l]$.
				\STATE \textbf{end for} 
				\STATE According to (\ref{PA}), obtain $\gamma_{\rm blk}$ and ${\bar{\bf x}}_{\rm pa}[l],\  \forall l \in {\mathcal{ L}}$.
				\STATE \textbf{for} $l=1$ \textbf{to} $L$ \textbf{do} 
				\STATE \quad \quad ${\bar P}_{\rm T}[l] = \left( \frac{\gamma_{\rm blk}}{\gamma[l]}\right)^2 P_{\rm T}[l]$, $\gamma[l]= \gamma_{\rm blk}$.
				\STATE \quad\quad\textbf{repmat}
				\STATE \quad\quad\quad Update ${\bar{\bf x}}_{\rm pa}[l]$ using the steps 5-8 in Algorithm \ref{A1}.
				\STATE \quad\quad\textbf{until} some stopping criterion is met.
				\STATE \textbf{end for}
				\STATE \textbf{Output:} $\gamma_{\rm blk}, {\bar{\bf x}}_{\rm pa}[l],\  \forall l \in {\mathcal{ L}}$.
			\end{algorithmic}\label{block transmission}
		\end{algorithm}
		
		By deploying the block transmission scheme at BS, ${\bar y}_{k}[l]$ is given by
		\begin{align}
			{\bar y}_{k}[l] = ({{\bf h}^T_k{\bf x}[l] + n_k[l]})/\gamma_{\rm blk}.
			\label{model after PA}
		\end{align}
	
	\section{Numerical Results}\label{results}
	In this section, we use the Monte Carlo method to evaluate the performance of the proposed methods in the scenario of an MU-MISO system and Rayleigh fading channel, with $P_{\rm T}[l]=1$ and ${\rm SNR}=\frac{1}{\sigma^2}$. We employ ZF and RZF precoding with symbol-level power constraints in \cite{Li2021}, CI-based SINR balancing (CISB) precoding \cite{Li2021}, and CI-based MMSE (CIMMSE) precoding \cite{9910472} as the baselines, with the final one chosen as the initial value for \textbf{Algorithm \ref{A1}}. To unify $\gamma$ of the entire block to avoid the excessive signaling overhead \cite{li2020symbol}, we employ \textbf{Algorithm \ref{block transmission}} for our proposed SLP, denoted by `ASM', and the PA scheme \eqref{PA} is used for ZF, RZF, CISB, and CIMMSE precodings.
	\begin{figure}[htp]
		\centering
		\subfigure[$K=N=8$.]{
			\includegraphics[width=3.2in]{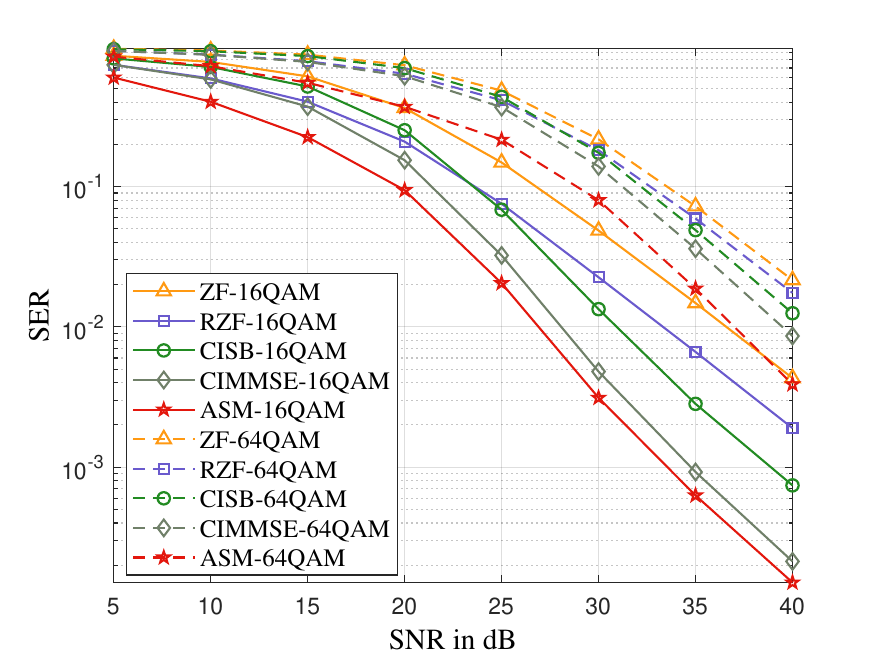}
			\label{block 8_8}
		}
		\subfigure[$K=N=12$.]{
			\includegraphics[width=3.2in]{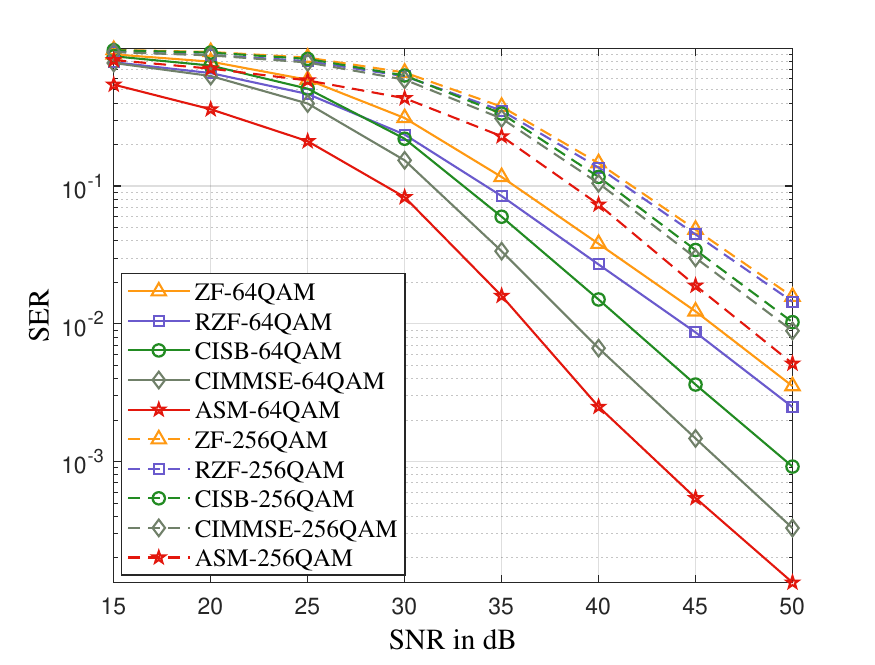}
			\label{block 12_12}
		}
		\DeclareGraphicsExtensions.
		\caption{SER vs SNR, $L=500$.}
	\end{figure}

	Fig. \ref{block 8_8} shows the comparison of methods for the $8\times 8$ MISO with 16QAM and 64QAM when $L=500$. It can be observed that, among the two typical scenarios, the SER performance of our proposed approach is lower than other methods over the full SNR range. With 16QAM and 64QAM, ASM provides SNR gains of about 1.2dB and 2.5dB over CIMMSE, respectively, when the SER performance is $10^{-2}$. Fig. \ref{block 12_12} compares the SER performance for the $12\times 12$ MISO.
	With 64QAM, ASM provides an SNR gain of about 2.5dB than CIMMSE when SER performance is $10^{-2}$, and the same trend is observed with 256QAM. It is noteworthy that our proposed approach provides more excellent performance when employed for higher-order QAM transmission.
	
\begin{figure}[hbtp]
		\centering
		\includegraphics[width=3.2in]{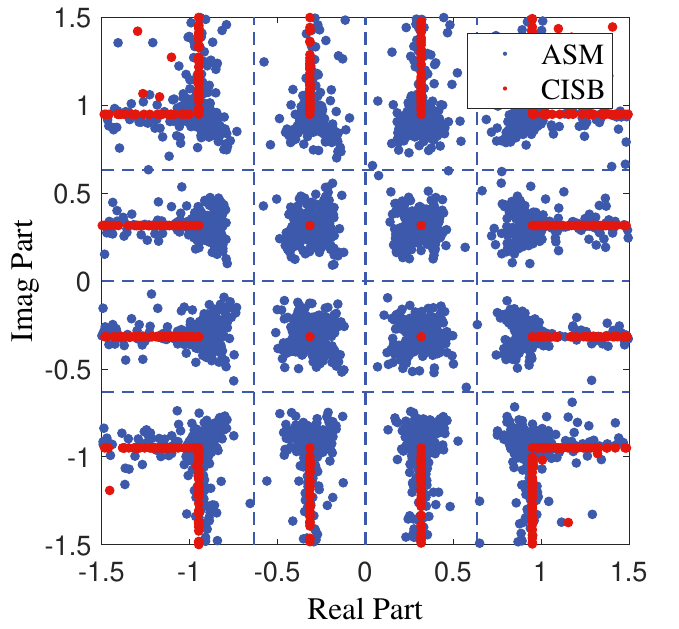}
		\caption{Scatterplot of noise-free received signals $\{({{\bf h}^T_k{\bf x}[l]})/\gamma_{\rm blk}\}_{k\in\mathcal{K},l\in\mathcal{L}}$ from ASM and CISB, 16QAM, $N=K=8$, $L=1024$, ${\rm SNR}=25$dB. }
		\label{Scatterplots}
	\end{figure}

	Scatterplot of noise-free received signals $\{({{\bf h}^T_k{\bf x}[l]})/\gamma_{\rm blk}\}_{k\in\mathcal{K},l\in\mathcal{L}}$ from ASM and CISB is depicted in Fig. \ref{Scatterplots}, where the dashed lines represent the decision boundaries. While CISB constrains these signals within the CIR \cite{Li2021}, ASM allows them to be freely distributed over the constellation map for lower transmission SER. Note that although the points from ASM are closer to the decision boundaries in Fig. \ref{Scatterplots}, $\gamma_{\rm blk}$ of ASM is much higher than that of CISB. This elevation in $\gamma_{\rm blk}$ reduces the influence of noise on the received signals from ASM and results in a considerably lower SER than the CISB signals. Besides, the irregular distribution of noise-free received signals from ASM reveals that the received signals corresponding to specific constellation points will follow non-Gaussian distributions.
	
	\section{Conclusion}\label{conclusion}
	In this paper, we investigated SLP for high-order QAM, aiming to minimize average SER. We first constructed the SER expression and put forward the ASM problem under a total transmit power constraint. To solve this non-convex problem, we proposed the DSAO algorithm, which alternatively minimizes the cost function on orthogonal Stiefel manifold and Euclidean spaces. To facilitate QAM demodulation in the receiver, we developed a block transmission scheme to keep the rescaling factor constant within a block. Simulation results demonstrated that the proposed SLP scheme exhibits a significant performance advantage over existing state-of-the-art SLP schemes.

	
	%

	\appendices
	\section{Gradient Expression in Algorithm \ref{A1}}\label{Gradient}
	$\nabla g({\bar{\bf x}})$ and $\nabla g(\gamma)$ are given by \eqref{gradient x} and \eqref{gradient gamma}, where $Q'(x)=-\frac{1}{\sqrt{2\pi}}e^{-\frac{x^2}{2}}$ and
	\begin{align}
	\delta^{{\rm R}^{+}}_k =\frac{\gamma {\hat s}^{\rm R}_k-{\bf f}^T_k{\bar {\bf x}}+\gamma d}{\sigma/\sqrt{2}},\ \delta^{{\rm R}^{-}}_k = \frac{\gamma {\hat s}^{\rm R}_k-{\bf f}^T_k{\bar {\bf x}}-\gamma d}{\sigma/\sqrt{2}},\\
	\delta^{{\rm I}^{+}}_k=\frac{\gamma {\hat s}^{\rm I}_k-{\bf g}^T_k{\bar {\bf x}}+\gamma d}{\sigma/\sqrt{2}},\ \delta^{{\rm I}^{-}}_k =\frac{\gamma {\hat s}^{\rm I}_k-{\bf g}^T_k{\bar {\bf x}}-\gamma d}{\sigma/\sqrt{2}},\\
	{\hat s}^{{\rm R}+}_k \!=\! d \!+\! {\hat s}^{{\rm R}}_k,{\hat s}^{{\rm R}-}_k \!=\! d \!-\!{\hat s}^{{\rm R}}_k,\ {\hat s}^{{\rm I}+}_k \!=\! d \!+\!{\hat s}^{{\rm I}}_k,{\hat s}^{{\rm I}-}_k \!=\! d \!-\!{\hat s}^{{\rm I}}_k.
	\end{align}
	
	\begin{figure*}[hb]
		\normalsize
		\setcounter{MYtempeqncnt}{\value{equation}}
		\vspace*{4pt}
		\hrulefill
		\begin{align}
		\nabla \! g({\bar{\bf x}})\!=\! \frac{\sqrt{2}}{\sigma}\sum_{k=1}^{K}\left(\left[Q^{'}(\delta^{{\rm R}^{-}}_k)\!-\! b^{\rm R}_kQ^{'}(\delta^{{\rm R}^{+}}_k)\right]\!\!
				\left[Q(\delta^{{\rm I}^{-}}_k)\!-\! b^{\rm I}_kQ(\delta^{{\rm I}^{+}}_k)\right]{\bf f}_k \!+\!
				\left[Q(\delta^{{\rm R}^{-}}_k)\!-\! b^{\rm R}_kQ(\delta^{{\rm R}^{+}}_k)\right]\!\!
				\left[Q^{'}(\delta^{{\rm I}^{-}}_k)\!-\! b^{\rm I}_kQ^{'}(\delta^{{\rm I}^{+}}_k)\right]{\bf g}_k\right).
		\label{gradient x}
		\end{align}
		\begin{align}
		\nabla \! g(\gamma) \!=\! \frac{\sqrt{2}}{\sigma}\!\sum_{k=1}^{K}\!\left(\left[{\hat s}^{{\rm R}^{-}}_kQ^{'}(\delta^{{\rm R}^{-}}_k)\!+\! b^{\rm R}_k{\hat s}^{{\rm R}^{+}}_kQ^{'}(\delta^{{\rm R}^{+}}_k)\right]\!\!
				\left[Q(\delta^{{\rm I}^{-}}_k)\!-\! b^{\rm I}_kQ(\delta^{{\rm I}^{+}}_k)\right]\!+\!
				\left[Q(\delta^{{\rm R}^{-}}_k)\!-\! b^{\rm R}_kQ(\delta^{{\rm R}^{+}}_k)\right]\!\!
				\left[{\hat s}^{{\rm I}-}_kQ^{'}(\delta^{{\rm I}^{-}}_k)\!+\! b^{\rm I}_k{\hat s}^{{\rm I}^{+}}_kQ^{'}(\delta^{{\rm I}^{+}}_k)\right]\right)\!.
		\label{gradient gamma}
		\end{align}
	\end{figure*}
	\ifCLASSOPTIONcaptionsoff
	\newpage
	\fi

	
	
	%
	
	\bibliography{IEEEfull}		  
	\bibliographystyle{IEEEtran}    
		
\end{document}